\documentclass[twocolumn]{aastex631}
\usepackage{amsmath}
\shorttitle{Ellipticities from HSSC}
\shortauthors{Liu et al.}
\graphicspath{{./}{figures/}}

\begin{document}

\title{Ellipticities of Galaxy Cluster Halos from Halo-Shear-Shear Correlations}

\correspondingauthor{Jun Zhang}
\email{betajzhang@sjtu.edu.cn}

\author[0009-0000-3649-5365]{Zhenjie Liu}
\affiliation{Department of Astronomy, Shanghai Jiao Tong University, Shanghai 200240, China}

\author{Jun Zhang}
\affiliation{Department of Astronomy, Shanghai Jiao Tong University, Shanghai 200240, China}
\affiliation{Shanghai Key Laboratory for Particle Physics and Cosmology, Shanghai 200240, China}

\author{Cong Liu}
\affiliation{Department of Astronomy, Shanghai Jiao Tong University, Shanghai 200240, China}

\author{Hekun Li}
\affiliation{Shanghai Astronomical Observatory, Chinese Academy of Sciences, Shanghai 200030, China}



\begin{abstract}
We report the first detection of the halo ellipticities of galaxy clusters by applying the halo-shear-shear correlations (HSSCs), without the necessity of major axis determination. We use the Fourier\_Quad shear catalog based on the Hyper Suprime-Cam Survey and the group catalog from the Dark Energy Spectroscopic Instrument Legacy Surveys for the measurement of group/cluster lensing and HSSC. Our analysis includes the off-centering effects. We obtain the average projected ellipticity of dark matter halos with mass $13.5 < {\rm log} (M_G h/ M_\odot) < 14.5$ within 1.3 virial radius to be $0.48^{+0.12}_{-0.19}$. We divide the sample into two groups based on mass and redshift, and we find that halos with higher mass tend to exhibit increased ellipticity. We also reveal that high-richness halos have larger ellipticities, confirming the physical picture from numerical simulation that high-richiness halos have a dynamical youth and more active mass accretion phase. 
\end{abstract}



\section{Introduction}

Dark matter halos are large-scale structures formed by the gravity of dark matter. They originate from tiny density fluctuations in the early universe, which collapse and cluster under the influence of gravity, and then evolve and form through continuous merging and accretion within the cosmic web \citep{dovich1970}. N-body simulations of dark matter reveal the triaxiality of dark matter halos, influenced by the direction of the last major merger and the accretion process along the filaments \citep{Lau2021}. Over time, this connectivity with the cosmic web weakens, and accretion becomes more isotropic, leading to a more spherical shape of halos \citep{kasun2005shapes, Allgood2006, suto2016evolution, Cataldi2023}. Particularly in the inner regions, baryonic matter, through processes such as star formation and energy feedback, drives the dark matter halos from triaxial to rounder shapes \citep{Illustris2019}. Various simulations also report that higher mass halos tend to be less spherical than lower mass halos \citep{jing2002triaxial,Allgood2006,Despali2014, Bonamigo2015}, mainly because high-mass halos form later and are more influenced by the surrounding filamentary structure, resulting in stronger triaxiality. Additionally, different dark matter models can affect the ellipticity of halos. For instance, Self-Interacting Dark Matter (SIDM) due to particle collisions tends to produce rounder halos, especially within the interiors of halos \citep{Peter2013,gonzalez2024cluster}.

A common method to measure ellipticity involves the gravitational lensing effect around dark matter halos. When light passes near a halo, it is bent by the gravitational field of the halo, resulting in distorted images of galaxies. This distortion, commonly referred to as shear, is primarily aligned in the direction tangential to the mass distribution of the lens \citep{Bartelmann2001}. There have been studies that measure the ellipticity of dark matter halos at the galaxy or cluster scale through strong lensing \citep{oguri2012combined, limousin2013three, Bruderer2015, Jauzac2018}, convergence map reconstruction \citep{Oguri2010}, and two-dimensional galaxy-galaxy lensing \citep{van2012constraints, clampitt2015lensing, schrabback2015cfhtlens, Clampitt2016, van2017, shin2018ellipticity, dvornik2019case, gonzalez2021measuring, schrabback2021tightening,Georgiou2021}. Strong gravitational lensing and convergence field analysis are limited to a small number of massive foreground samples and the inner radial regions. Galaxy-galaxy lensing, which is the cross-correlation between foreground positions and background shear, can analyze the average properties of a large number of low-mass halos by stacking the samples, especially stacking along the major axis to study their anisotropy. However, the actual major axis orientation of halos is unknown, and it is often assumed that the direction of the brightest cluster galaxy (BCG) or satellite galaxies aligns with the halo. This method faces the issue of misalignment of the major axis \citep{Jauzac2018, Okabe2020}, which would dilute the signal and result in bias.

\cite{simon2012towards} shows that the third-order galaxy-galaxy lensing is sensitive to the ellipticity of dark matter halos and their substructures. Subsequently, \cite{Adhikari2015JCAP} proposes the use of halo-shear-shear correlation (HSSC) to measure the projected ellipticity of halos to avoid the biases in ellipticity measurements caused by galaxy-halo misalignments. HSSC is a correlation between the shears of two background galaxies around a halo, which reflects the morphology of the halo. They develop an estimator for the halo ellipticity using a simple model of the projected surface density profile of halos and validate it with simulations. Furthermore, \cite{HSSC2017} conduct a detailed study on the effects of substructures, projection effects, and off-centering on HSSC measurements using N-body simulations. 

In our work, for the first time, we measure the HSSC of halos in galaxy clusters using the observational shear data. It is mainly for the purpose of constraining the average ellipticity for a large sample of cluster halos.
Our analysis includes the off-centering effect. 
\S\ref{sec:model} illustrates the theoretical model, including models of HSSC and off-centering effects. The data and the method of measurements are introduced in \S\ref{sec:data} and \S\ref{sec:measure} respectively. We present our main results in \S\ref{sec:results}, and provide comparison with the results from hydrodynamic simulations for discussions.
Finally, we conclude in \S\ref{sec:conclusion}.

\section{Model}\label{sec:model}
\subsection{Multipole Approximation}
Due to the evolutionary history and environmental influences, halos inevitably exhibit anisotropy, such as a tendency to elongate along the direction of matter inflow or the filamentary structures. When considering the anisotropy of halos, their density field can be decomposed into monopole and quadrupole terms. The monopole term represents the average density of the halo, while the quadrupole term captures the deviation from sphericity and describes the shape of the halo. Typically, the ellipticity of halos is deemed to be quite small. Following the derivation in \cite{van2017}, the ellipticity is assumed to be minor, allowing the multipole expansion of the surface density to be succinctly expressed as
\begin{equation}
\Sigma(R)=\Sigma(r, \theta) \cong \Sigma_0(r)+\varepsilon \Sigma_2(r) \cos 2 \theta
\end{equation}
where $R^2=r^2\left(q \cos ^2 \theta+\frac{\sin ^2 \theta}{q}\right)$, $\theta$ represents the angle between the major axis and background galaxy and $\Sigma_2=-r \frac{d \Sigma_0(r)}{d r}$. $\varepsilon$ is the projected ellipticity, related to the axis ratio $q$ by \begin{equation}\label{eq:eq}
\varepsilon=(1-q) /(1+q).
\end{equation}
Moreover, we can assume that the shear components takes the following form:
\begin{eqnarray}
\label{eq:gt}
&&\gamma_t(r, \theta)=\gamma_{t0}(r)+\varepsilon \gamma_{t2}(r) \cos 2 \theta \\ \nonumber
&&\gamma_\times(r, \theta)=\varepsilon \gamma_{\times 2}(r) \sin 2 \theta
\end{eqnarray}
When neglecting the orientation of dark matter halos, the second-order terms of $\gamma_{t/\times}$ will be averaged out, thus the measurement results will revert to the traditional stacked galaxy-galaxy lensing, which includes only the isotropic zero-order components, $\gamma_{t0}$ for $\gamma_t$ and 0 for $\gamma_\times$. The relationship between $\gamma_{t0}$ and surface density is 
\begin{equation}\label{eq:esd}
\Sigma_c \gamma_{t0}(r)=\Delta \Sigma (r)\equiv\overline{\Sigma}_0(<r)-\Sigma_0(r),
\end{equation}
where $\overline{\Sigma}_0(<r)$ refers to the average surface density within a radius $r$, and we term $\Delta \Sigma (r)$ the excess surface density (ESD). $\Sigma_c$ is the comoving critical surface density, defined as 
\begin{equation}\label{sigmac}
    \Sigma_c=c^2 D_{\rm s}/[4 \pi G D_{\rm l} D_{\rm l s}(1+z_{\rm l})^2]
\end{equation} 
Here, $c$ is the speed of light, $G$ is the gravitational constant, and $D_{\rm s}, D_{\rm l}$, and $D_{\rm ls}$ are the angular diameter distances for the lens, source, and lens-source systems, respectively. We employ the analytic formula of the surface density for the Navarro–Frenk–White (NFW, \cite{navarro1997universal}) profile in \cite{yang2006} to describe the isotropic part of halos, i.e. $\Sigma_0(r)$. In the following text, for convenience, we denote $\Sigma_c \gamma_i$ as $\Gamma_i$, so that $\Delta \Sigma=\Gamma_{t0}$.

For the second-order terms of shears, by solving the Poisson equation, the formula for the second-order term of $\gamma_t$ can be derived,
\begin{equation}
\Gamma_{t2}(r)\equiv\Sigma_c \gamma_{t2}(r)=-\frac{6 \psi_2(r)}{r^2}-2 \Sigma_0(r)-\Sigma_2(r)
\end{equation}
\begin{equation}
\Gamma_{\times 2}(r)\equiv\Sigma_c \gamma_{\times 2}(r)=-\frac{6 \psi_2(r)}{r^2}-4 \Sigma_0(r)
\end{equation}
where $\psi_2$ is the quadrupole component of the lensing potential given by: 
\begin{equation}
\psi_2=-\frac{2}{r^2} \int_0^r r^{\prime 3} \Sigma_0\left(r^{\prime}\right) d r^{\prime}.
\end{equation}
In this way, we can obtain the multipole moments of the shear field by solely employing a spherically symmetric halo density distribution model.

The surface density of an ellipsoidal halo varies with angle, leading to corresponding variations in shear, as shown in Eq.\ref{eq:gt}. Naturally, by correlating shears on different angle around a halo, the effect from the second-order ellipticity-dependent terms emerge, eliminating the need to define the axis direction. Ignoring other factors like off-centering and substructure, we can straightforwardly multiply the tangential shears from varying orientations as follows,
\begin{equation}\label{xittcen}
\begin{aligned}
 &\zeta_{\rm tt}^{\rm cen}(r_1,r_2,\beta) \\
 &= \left\langle \Gamma_t(r_1,\theta) \Gamma_t \left(r_2,\theta+\beta\right)\right\rangle_\theta \\
&=  \Gamma_{t0}(r_1) \Gamma_{t0}(r_2) + \frac{\varepsilon^2}{2}  \Gamma_{t2}(r_1) \Gamma_{t2}(r_2) \cos 2 \beta .
\end{aligned}
\end{equation}
where $\beta$ denotes the angular difference of two background galaxies relative to the center. It is evident that this shear correlation does not depend on the orientation of the major axis, 
and the amplitude of its angular variation depends on the ellipticity of the halo. Similarly, the cross component also has nonzero correlations with itself or the tangential one as follows:
\begin{eqnarray}\label{xixxcen}
\zeta_{\rm \times \times}^{\rm cen}(r_1,r_2, \beta) &=& \frac{\varepsilon^2}{2}  \Gamma_{\times 2} (r_1) \Gamma_{\times 2}(r_2) \cos 2 \beta \\ \nonumber
\zeta_{\rm t \times}^{\rm cen}(r_1,r_2, \beta) &=& \frac{\varepsilon^2}{2}  \Gamma_{t 2} (r_1) \Gamma_{\times 2}(r_2) \sin 2 \beta \\ \nonumber
&=&  -\zeta_{\rm \times t}^{\rm cen}(r_1,r_2, \beta).
\end{eqnarray}

\subsection{Off-centering Model}\label{sec:off}
In practice, inaccuracy in the halo center complicates both the modeling and measurement processes. Here, we discuss two aspects of our off-centering model: 1. the impact on the shear signals and HSSCs; 2. the anisotropy of the distribution of the true halo center with respect to the chosen center.

\begin{figure}[ht!]
\centering
\includegraphics[width=0.4\textwidth]{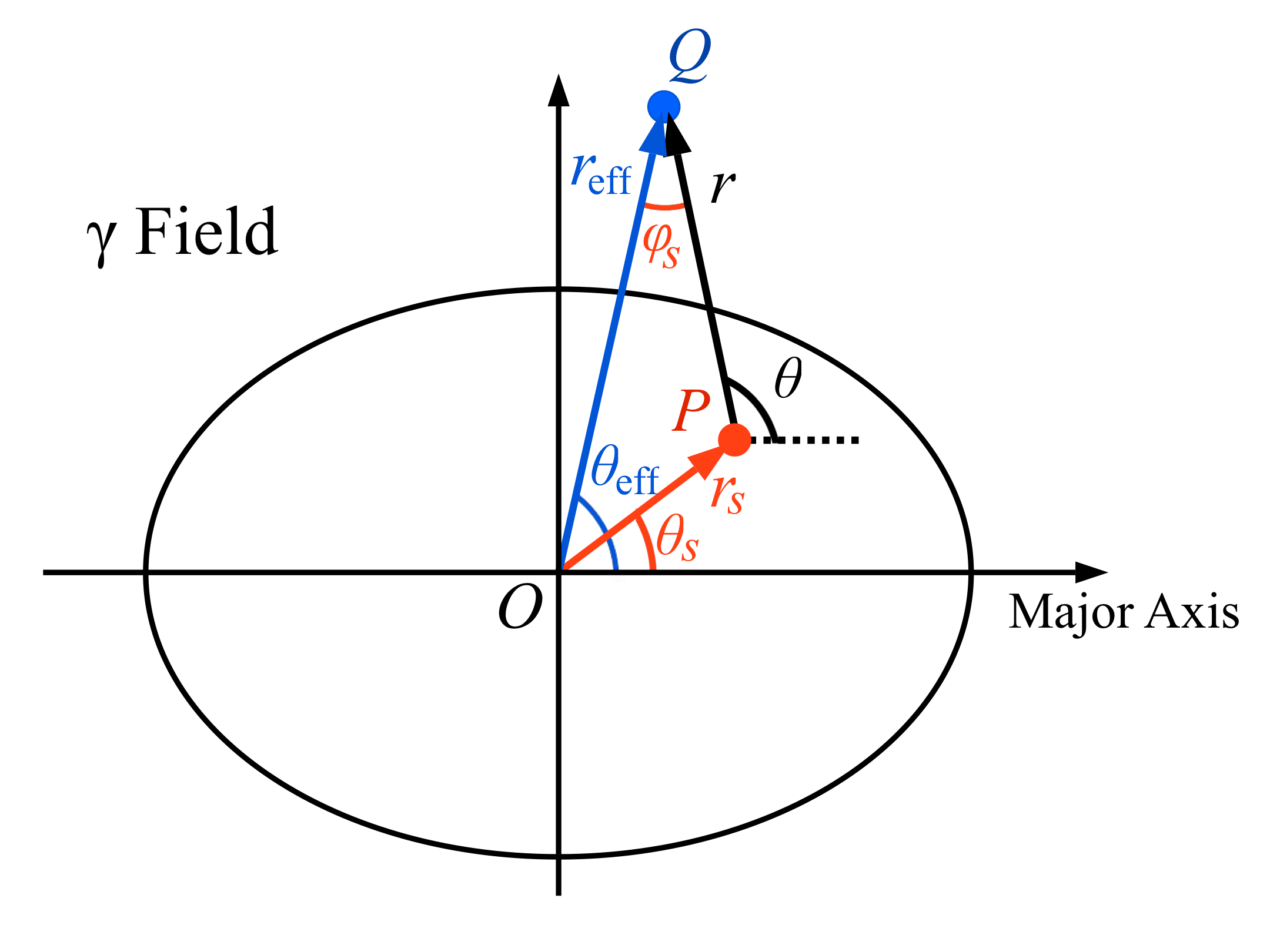}
\caption{Schematic diagram of off-centering. The plane displays the shear field of the halo, where the ellipse represents the dark matter halo, with its major axis aligned along the x-axis, and $O$ is the halo center set as the coordinate origin. $P$ is the incorrect halo center, with $(r_s,\theta_s)$ being its position relative to $O$. $Q$ is the location of the shear measurement, with coordinates $(r_{\rm eff},\theta_{\rm eff})$ relative to $O$. The angle $\varphi_s$ is between the true halo center and the incorrect center. \label{fig:dev}}
\end{figure}

\subsubsection{Off-centering effect for single halo}
In lensing measurements, off-centering not only results in the misidentification of background galaxies but also leads to the erroneous rotation direction of shear estimators. The shear fields of $\gamma_{t / \times}$ in Eq.\ref{eq:gt} are relative to halo center. But in observations, we rotate the shear towards the assumed centers, such as BCGs or luminosity-weighted centers (LWCs), which would lead to some biases of observed shears. Hence, we should model the effect of off-centers. As depicted in the shear field surrounding a halo in Figure \ref{fig:dev}, with the major axis on the horizontal. The point $O$ represents the halo center and coordinate center, the point $P$ signifies the incorrectly assumed center at $(r_s,\theta_s)$, and $Q$ marks the location of a background galaxy at $(r_{\text {eff }}, \theta_{\text {eff}})$. The observed shear components $\gamma_{t / \times}^{\rm obs}$ at $(r,\theta)$ defined with respect to the point $P$ can be rotated from the components $\gamma_{t / \times} \left(r_{\text {eff }}, \theta_{\text {eff}}\right)$ relative to $O$ via the following form:
\begin{equation}\label{eq:gtobs}
\left[\begin{array}{l}
\gamma_t^{\text {obs }} \\
\gamma_\times^{\text {obs }}
\end{array}\right]=\left[\begin{array}{cc}
\cos 2 \varphi_s & \sin 2 \varphi_s \\
-\sin 2 \varphi_s & \cos 2 \varphi_s
\end{array}\right]\left[\begin{array}{l}
\gamma_t(r_{\text{eff}}, \theta_{\text{eff}}) \\
\gamma_\times (r_{\text{eff}}, \theta_{\text{eff}})
\end{array}\right]
\end{equation}
where $\varphi_s$ is the angle between the true halo center and the incorrect center. For given values of $r$, $\theta$, $r_s$, and $\theta_s$, we can solve for $r_{\text{eff}}$, $\theta_{\text{eff}}$, and $\varphi_s$ using the following relations:
\begin{eqnarray}
 &&r_{\text {eff }}=\sqrt{r^2+r_s^2+2 r_s r \cos \left(\theta-\theta_s\right)}, \\ \nonumber
&&\cos \left(\theta_{\text {eff }}-\theta_s\right)=\frac{r_s^2+r_{\text {eff}}^2-r^2}{2 r_s r_{\text {eff}}},\\ \nonumber
&&\cos (\varphi_s)=\frac{r^2+r_{\text {eff}}^2-r_s^2}{2 r r_{\text {eff}}}, \\ \nonumber
&&\frac{r}{\sin \left(\theta_{\text {eff }}-\theta_s\right)}=\frac{r_{\text {eff }}}{\sin \left(\pi-\theta+\theta_s\right)}=\frac{r_s}{\sin(\varphi_s)}.
\end{eqnarray}
Therefore, for a single halo with a given offset center $(r_s, \theta_s)$, its observed HSSCs should be modeled as
\begin{equation}
\begin{aligned}\label{eq:obsij}
 \zeta_{\rm ij}^{\rm obs} \left(r_1,r_2,\beta , r_s, \theta_s \right)&=\frac{1}{2 \pi} \int_0^{2 \pi} {\rm d} \theta \Gamma_i^{\text{obs}}(r_1, \theta , r_s, \theta_s ) \\
 &\times \Gamma_j^{\text{obs}}\left(r_2, \theta+\beta , r_s,  \theta_s \right)
\end{aligned}
\end{equation}
where $\Gamma_{i/j}^{\text{obs}}$ represents $\gamma_{i/j}^{\text{obs}}$ in Eq.\ref{eq:gtobs} multiplied by $\Sigma_c$. In \S\ref{app:off}, we investigate the impact of various off-centers on the HSSCs in our theoretical model. Our analysis reveals that the presence of an off-center introduces asymmetry in the shapes of $\zeta_{\rm tt}(\beta)$ and $\zeta_{\times \times}(\beta)$, and results in a suppression of $\zeta_{\rm tt}$, particularly for smaller radii. Additionally, the off-centers exert a larger influence on components involving $\gamma_\times$. For detailed information, please refer to \S\ref{app:off}.



\subsubsection{The Distribution of Off-centers}

\begin{figure}[ht!]
\centering
\includegraphics[width=0.4\textwidth]{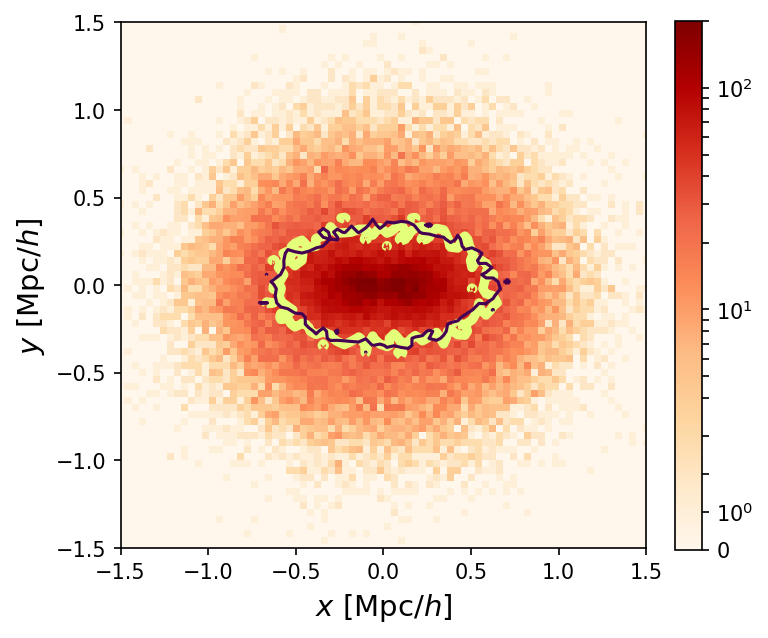}
\caption{The projected distribution of the stacked satellite galaxies (red), centered around the luminosity-weighted center (LWC) of each cluster and aligned with the long axis ($x$-axis). The yellow contour is an isocontour of the satellite distribution (red). The black isocontour represents the distribution of BCGs relative to LWCs, encompassing about 97\% of the BCG positions. }\label{fig:BCGs}
\end{figure}

In the monopole model of galaxy-galaxy lensing, we only need to consider the radial distribution of off-centering, and a Rayleigh distribution \citep{offcen2007} is considered to be the most realistic,
\begin{equation}\label{offr}
P_r\left(r_s \right)=\frac{r_s }{\sigma_s^2} \exp \left(- \frac{r_s^2}{\sigma_s^2}\right),
\end{equation}
where $\sigma _s$ describes the dispersion of the center's deviation. 
However, considering the anisotropic part of halos, the angular distribution of off-centering $P\left(r_s, \theta_s \right)$ could become important, as it exhibits a certain degeneracy with the halo ellipticity. For modeling $P\left(r_s, \theta_s \right)$ of BCGs, we show the distribution of stacked satellite galaxies (red points) and BCGs (contours) from the cluster catalog (described in \S\ref{lencat}) in Figure \ref{fig:BCGs}, using the luminosity-weighted centers. The $x$-axis represents the major axis direction of the satellite galaxies, and the direction angle $\phi$ is determined by the quadrupole moments:
\begin{equation}
{\rm tan} 2 \phi =\frac{2 Q_{12} }{Q_{11}-Q_{22}}.
\end{equation}
\begin{equation}\label{Qij}
Q_{i j} = \frac{\sum_k(x_{i, k} x_{j, k} w_k) }{ \sum_k w_k},
\end{equation}
where $(x_{i, k} x_{j, k})$ are the coordinates of the $k^{\rm th}$ galaxy in the $i$ and $j$ directions relative to the luminosity weighted center, and the weight is $w_k=1 /(x_{1, k}^2+x_{2, k}^2)$. In Figure \ref{fig:BCGs}, the yellow line is an isocontour of the satellite galaxy distribution (red), while the black line is drawn in a nearby location to represent the isocontour of the BCGs. We notice that the angular distribution of the BCGs (black) is consistent with that of the satellite galaxies (yellow), so we assume that the distribution of BCGs relative to the true dark matter halo center has the same projected ellipticity $\varepsilon$ as the halo. Therefore, we obtain an anisotropic distribution of off-centers by simply transforming the coordinates in Eq.\ref{offr},
\begin{equation}\label{offrthe}
P\left(r_s, \theta_s \right)=P_r \bigg(r_s \sqrt{q(\varepsilon) {\rm cos}^2 (\theta_s) +{\rm sin}^2 (\theta_s) / q(\varepsilon)} \bigg)
\end{equation}
where $q(\varepsilon)$ is the inverse function of Eq.\ref{eq:eq}. Note that $q(\varepsilon)$ here represents the anisotropic distribution of off-centers, which could also be set to other values. In \S\ref{sec:offcen-result}, we discuss the impacts of off-centering model using two extreme cases of off-centering distributions, and find an insignificant effect in ellipticity measurements.

Taking into account the off-centering effect, the ESD can be modeled as \begin{equation}
\begin{aligned}
 \Delta \Sigma^{\rm off}\left(r \right)&=\frac{1}{4 \pi^2} \int_0^{\infty} {\rm d} r_s \int_0^{2 \pi} {\rm d} \theta_s P\left(r_s, \theta_s \right)   \\
&\times \int_0^{2 \pi} {\rm d} \theta \Gamma_t^{\text{obs}}(r,\theta , r_s, \theta_s )  ,
\end{aligned}
\end{equation}
and the HSSC model can be formulated as
\begin{equation}
\begin{aligned}\label{eq:offtt}
 \zeta_{\rm tt}^{\rm off} \left(r_1,r_2,\beta \right)&=\frac{1}{4 \pi^2} \int_0^{\infty} {\rm d} r_s \int_0^{2 \pi} {\rm d} \theta_s P\left(r_s, \theta_s \right)  \\
& \times \zeta_{\rm tt}^{\rm obs} \left(r_1,r_2,\beta,r_s, \theta_s \right)
\end{aligned}
\end{equation}
where $\zeta_{\rm tt}^{\rm obs}$ is expressed in Eq.\ref{eq:obsij}. Ultimately, we use the parameter $f_{\rm c}$ to depict the proportion of well-centered clusters, so the final ESD and HSSC can be modeled as:
\begin{equation}\label{dsigmatot}
\Delta \Sigma (r)=f_{\rm c}\Delta \Sigma_0(r) +(1-f_{\rm c})\Delta \Sigma^{\rm off}\left(r \right) 
\end{equation}
and
\begin{equation}
\begin{aligned}\label{zetatot}
\zeta_{\rm tt}(\beta)&=\int_{r_{\rm min}}^{r_{\rm max}} {\rm d} r_1 \int_{r_{\rm min}}^{r_{\rm max}} {\rm d} r_2 P_{\rm b}(r_1,r_2) \\
&\times \big[f_{\rm c}\zeta_{\rm tt}^{\rm cen}(r_1,r_2, \beta) +(1-f_{\rm c})\zeta_{\rm tt}^{\rm off}\left(r_1,r_2,\beta \right) \big] .
\end{aligned}
\end{equation}
where $r_{\rm min}$ and $r_{\rm max}$ respectively refer to the minimum and maximum radii in HSSC measurements, and $P_{\rm b}(r_1,r_2)$ is the probability of background galaxies in different radius pairs. Additionally, our theoretical framework necessitates integrating the ESD in Eq.\ref{dsigmatot} and the HSSC in Eq.\ref{zetatot} over the mass and redshift distribution of the lens sample to match observed measurements, and they are also weighted by the number of background galaxies. 

\section{Data}\label{sec:data}

\subsection{Shear Catalog}
 In our analysis, we utilize data from the third public data release of Hyper Suprime-Cam (HSC) Survey \citep{Aihara2022}, which is known for its deep optical observations of the universe, reaching a limiting magnitude of about 26 with excellent spatial resolution. The images have five bands: $g$, $r$, $i$, $z$, and $y$, with the $i$ band providing the highest imaging quality. Our shear catalog is processed through the Fourier\_Quad (FQ) pipeline \citep{Liu2024zlh}, covering around 1400 deg$^2$ and containing about 100 million galaxies. This catalog includes the basic information of galaxies such as 3D position, signal-to-noise ratio $\nu_F$ \citep{Li2021}, magnitude, and shear estimates.  The FQ shear estimators are derived from the multipole components of galaxy power spectrum, comprising five estimators: $G_1$, $G_2$, $N$, $U$, and $V$, where $G_i$ resembles the ellipticity components $e_i$, and $N$ serves as a normalization factor, $U$ and $V$ are additional correction terms, detailed in \cite{Zhang2017}.
 
 To enhance measurement accuracy, we select galaxies with $\nu_F > 4$. Photometric redshifts (photo-$z$) are obtained using the DEmP method from \cite{Nishizawa2020}, also providing their uncertainty estimates $\sigma_z$. From Figure 4 in \cite{Nishizawa2020}, it can be observed that at photo-$z > 1.5$, the redshift bias and scatter become significant. Therefore, to mitigate the bias caused by photo-$z$ errors and outliers, galaxies with $\sigma_z > 0.05$ are excluded from our analysis, leading to almost no galaxies with $z > 1.5$ in the final source sample. In our lensing analysis, we ensure background galaxies had a photo-$z$ greater than the lens redshift by 0.2, i.e., $z_s > z_l + 0.2$, to reduce the dilution of cluster and background galaxies and intrinsic alignment contamination. Additionally, the FQ pipeline performs shear measurements on galaxy images for each exposure in different bands, and the shear correlations on the same exposure may be affected by some systematics, e.g., correlated PSF residuals \citep{Lu_2018}. Thus, we select galaxy pairs from different exposures for the HSSC measurement to avoid potential biases. Based on \cite{Liu2024zlh}, using data from multiple bands can improve the signal-to-noise ratio in lensing, hence we chose data from the $r$, $i$, and $z$ bands as our shear samples. 
 
\subsection{Lens Catalog}\label{lencat}
Our galaxy cluster catalog is from \cite{yang2021extended}, which is based on the Dark Energy Spectroscopic Instrument \citep[DESI;][]{2019AJ....157..168D} Image Legacy Surveys DR9. This catalog provides detailed information on the 3D coordinates, richness $\lambda$, halo mass ${\rm log}M_G$ and total galaxy luminosity of the clusters. The halo mass estimated using the abundance matching method, ranging from $10^{11.5} M_\odot/h$ to $10^{14.5} M_\odot/h$, with an uncertainty of about 0.2 dex at the high-mass end ($M_G > 10^{13.5} M_\odot /h$) and about 0.45 dex at the low-mass end.  In our research, to ensure a high signal-to-noise ratio, we focus on clusters with masses between $10^{13.5}-10^{14.5} M_\odot/h$ without richness selection. Additionally, to ensure a sufficient number of background galaxies, we select clusters with $0.2 < z_l < 0.5$. Given the relatively small footprint of HSC, we filter our sample to include clusters that overlap with the HSC fields. Ultimately, we identify 23298 lens samples. Notably, while the catalog provides luminosity weighted center of clusters, following \cite{Wang2022}, we use the position of the BCG as the center of the galaxy cluster for a more accurate tracking of the halo center in our main analysis.

\section{Measurement}\label{sec:measure}
In our work, we employ the PDF-symmetrization (PDF-SYM) method to measure the ESD and HSSC signals of galaxy clusters \citep{Zhang2017,Wang2022,Liu2023}. This approach is designed to maximize the statistical information from shear estimators and mitigate statistical biases due to uneven or finite distributions of background sources. The essence of this method involves constructing the probability distribution function (PDF) for the shear signal (or the joint PDF for shear-shear correlations), followed by an optimization process to determine the ideal ESD or HSSC values that achieves a symmetric state of the PDF (or the joint PDF). Here, we provide a detailed introduction to the measurements of ESD and HSSC. Besides, we find that the components of HSSCs including $\gamma_\times$ is significantly impacted by off-centering effects and they show low signal-to-noise ratios due to the opposite sign at different radius as shown in \S\ref{app:off} and \S \ref{sec:all}. Therefore, we rely solely on the results from the $\gamma_t$ component for parameter constraints. Before performing the lensing measurements, we test the bias in shear measurements using field distortion and find that the multiplicative biases are at most on the order of 1-2 percent (see \S\ref{sec:fd}). Meanwhile, the additive bias is expected to be subtracted using the corresponding statistics for random points.

\subsection{ESD}
In applying the PDF-SYM method, we first stack the tangential shear estimators $G_t$ of all background galaxies of lens samples into a PDF. Based on \cite{Zhang2017}, the unlensed shear estimators $G_t^{\rm S}=G_t-g_t (N+U_t)$, where the $g_t$ is the real shear of galaxies. The PDF of $G_t^{\rm S}$ is assumed to be symmetric with respect to 0, since background galaxies should be randomly oriented. Therefore, by varying the value of $\widehat{\Gamma}_t$, we could have different PDFs of
\begin{equation}\label{Ghat}
\begin{aligned}
\hat{G}_t & =G_t-\frac{\widehat{\Gamma}_t}{\Sigma_c}\left(N + U_t\right)\\
&= G_t^{\rm S} + \frac{\Sigma_c g_t -\widehat{\Gamma}_t}{\Sigma_c}\left(N + U_t\right).
\end{aligned}
\end{equation}
This step is trying to recover $G_t$ to its unlensed state whose PDF is symmetric. We achieve the most symmetric PDF of $\hat{G_t}$ by minimizing the $\chi^2$ (defined in Eq.36 of \cite{Zhang2017}). When stacking background galaxies without considering orientation, the optimal value obtained will be average ESD value of foreground masses in Eq.\ref{eq:esd}.  The feasibility and robustness of our method have been demonstrated in \cite{Wang2022}, employing the PDF-SYM technique with the shear catalog generated from DECaLS \citep{Zhang2022} and studying the halo properties and mass function of galaxy clusters \citep{yang2021extended}. To mitigate the significant impacts of baryon effects and other small-scale systematics, we set the minimum radius for all tests at 0.1 $r_{\rm vir}$, where $r_{\rm vir}$ is the virial radius defined from halo mass $M_G$ in the lens catalog. For the ESD measurements, 8 points were uniformly spaced on a log-scale from 0.1 to 2 $r_{\rm vir}$. Fitting $\Delta \Sigma$ allows us to glean more information about off-centering, which helps break the degeneracy with halo ellipticity in HSSC. 

\subsection{HSSC}
Furthermore, \cite{Zhang2017} introduce symmetrizing the joint PDF of shear estimators for measuring shear-shear correlations, and \cite{Liu2023} successfully apply it in cosmological analysis. Here, we apply the two-point statistics PDF-SYM method to measure the correlation between tangential shears around halos. Initially, we calculate the azimuthal angle for each background galaxy with respect to the BCG center, and identify galaxy pairs with an angular separation of $\beta$. In our measurements, since pairs separated by $\beta$ and $2\pi - \beta$ are duplicates, we uniformly divide $\beta$ into eight bins from 0 to $\pi$. We then stack the shear estimates $\hat{G}_t$ and $\hat{G}_t^\prime$ (defined in Eq. \ref{Ghat}) for all pairs into the joint PDF $P(\hat{G}_t, \hat{G}_t^\prime)$, assuming two sets of $\widehat{\Gamma}_t$ and $\widehat{\Gamma}_t^\prime$ for galaxy pairs in Eq. \ref{Ghat} with a covariance of the estimated value of HSSC $\widehat{\zeta_{\rm tt}}$, i.e., $\widehat{\zeta_{\rm tt}}=\langle \widehat{\Gamma}_t \widehat{\Gamma}_t^\prime \rangle$. As the estimated HSSC is exactly opposit to real HSSC, $\widehat{\zeta_{\rm tt}}=-\langle \Gamma_t \Gamma_t^\prime \rangle$, $P(\hat{G}_t, \hat{G}_t^\prime)$ becomes symmetrical. For details on how to generate two sets of $\widehat{\Gamma}_t$ and $\widehat{\Gamma}_t^\prime$, refer to the methods section in \cite{Liu2023}.

In two-point statistics, the shape of the observed field can also produce spurious signals, which need to be subtracted. Therefore, in our ESD measurements, we subtract the lensing results around random points at the same scales. Systematic errors caused by the shape of the field also exist in HSSC measurements. Besides, beyond individual dark matter halos, large-scale structure along the line of sight, including filaments, sheets and voids, also contribute to the shear correlation between two galaxies. This contamination is generally called the projection effect. Also, intrinsic alignment between two background galaxies can also lead to additional correlation signals. To remove the influence of projection effect and systematics from intrinsic alignment, we subtract the contribution from random points, i.e., random-shear-shear correlations, by
\begin{equation}\label{eq:rand}
\zeta_{ij}(r_1,r_2,\beta)=\zeta_{ij}(r_1,r_2,\beta;{\rm Halo})-\zeta_{ij}(r_1,r_2,\beta;{\rm Rand}).
\end{equation}
$\zeta_{ij}({\rm Halo})$ represents the HSSC around the halos in the measurement, centered on BCGs or LWCs. While $\zeta_{ij}({\rm Rand})$ represents the HSSC signal around random points, which is caused by the cosmological shear correlation of background galaxies and some systematics. The number of random points is about 10 times that of lenses in both ESD and HSSC measurements. Since the virial radius of each halo is defined to find background galaxies, these random points are assigned the same distributions of redshift and halo mass as the lens sample to reconstruct the additional cosmologic signal in $\zeta_{ij}({\rm Halo})$.

\subsection{Parameter Fitting}

\begin{table} 
\centering  
\caption{The prior of all fitting parameters in MCMC approach.} \label{tab:prior}
\begin{tabular}{l|cccc}
\hline  
 Parameter & $\varepsilon$ &$\alpha$& $\sigma_s$ &$f_{\rm c}$ \\
[2.4pt]
\hline
Prior & [0,1] &[0.5,1.5] & [0.1,1.5] & [0,1]  \\
\hline
\end{tabular}  
\end{table} 

Our analysis incorporates 4 free parameters: halo ellipticity $\varepsilon$, halo mass bias $\alpha$, well-centered proportion $f_{\rm c}$, and $\sigma_s$. The halo mass used is taken from the group catalog, but it is obtained from abundance matching and it could be problematic. We multiply the group mass by a factor $\alpha$ to represent the deviation in the halo mass, so the measured halo mass is $M_{\rm 180}=\alpha M_G$. We adopt the mass-concentration relation from \cite{duffy2008dark},
\begin{equation}
    c_{\rm 200}=5.71 \big( M_{\rm 200}/2 \times 10^{12} h^{-1} \big)^{-0.084}(1+z_l)^{-0.47}
\end{equation}
where $M_{\rm 200}$ and $c_{\rm 200}$
are defined based on a halo that is 200 times the mean matter density. To align with the halo mass definitions in the lens catalog, we used COLOSSUS \citep{diemer2018colossus} to convert these values to $M_{\rm 180}$ and $c_{\rm 180}$. The sky is divided into 100 subregions using the K-means algorithm in Scikit-learn \citep{scikit-learn}, with covariance matrix estimated via the Jackknife method and parameters constrained using the MCMC technique \citep{mcmc2000}. Due to the degeneracy of the mass parameter $\alpha$ with other parameters, we use both the ESD and the HSSC data within radius ranges $0.1 < r/r_{\rm vir} <0.5$ and $0.1 < r/r_{\rm vir} < 1.3$ to jointly constrain these four parameters, where $r_{\rm vir}$ is derived by the group mass $M_G$ in the catalog. The priors of the parameters in the fitting process are listed in Table \ref{tab:prior}.

\section{Results and discussion }\label{sec:results}
In this section, we present results of the measurements and the average ellipticities. We show the analysis of the projected ellipticity for all samples and vary the radial range of HSSC measurements to find changes in ellipticity within different radial regions. We assess the off-centering effect by comparing the results centered on BCGs and LWCs, and also examine different off-centering distribution models. Then, we test the dependency of ellipticity on properties such as mass, redshift and cluster richness by dividing the samples into two groups in each case. 

\begin{table*} 
\centering  
\caption{The constraints of four parameters at different radius range for different samples. $\alpha$ describes the bias of halo mass in the group catalog. The two different richness sub-samples have the same redshift and group mass distributions. The data points for ``All'' samples are shown in Figure \ref{fig:totres}, and those for other samples are displayed in Figure \ref{fig:highlow}.} \label{tab:baseres}
\begin{tabular}{l|ccc|ccccc}
\toprule
 Samples & log $M_G$ & $z$ & $N_{\rm lens}$& $\varepsilon$ & $\alpha$ &$\sigma_s$ & $f_c$ & $\chi^2$/d.o.f. \\
[2.4pt]
\hline
All   & &   & &$0.48^{+0.12}_{-0.19}$ & $0.84^{+0.04}_{-0.04}$ & $0.33^{+0.05}_{-0.04}$ & $0.59^{+0.03}_{-0.03}$ &1.03\\
 [2.4pt]
 All (inner)  & [13.5, 14.5) &  [0.2, 0.5) & 23298  &$0.41^{+0.18}_{-0.24}$ & $0.87^{+0.05}_{-0.04}$ & $0.37^{+0.07}_{-0.05}$ & $0.63^{+0.03}_{-0.03}$&1.65 \\
 [2.4pt]
 All (LWC)  & &   & &$0.17^{+0.14}_{-0.12}$ & $0.88^{+0.03}_{-0.03}$ & $0.17^{+0.01}_{-0.01}$ & $0.10^{+0.04}_{-0.05}$&1.59\\
 [2.4pt]
\hline
Low $M_h$   & [13.5, 14) &  [0.2, 0.5)  &  19905 &$0.36^{+0.23}_{-0.23}$ & $0.81^{+0.05}_{-0.05}$ & $0.33^{+0.07}_{-0.06}$ & $0.60^{+0.03}_{-0.04}$ &1.66\\
 [2.4pt]
High $M_h$   & [14, 14.5) &  [0.2, 0.5) &3393&$0.53^{+0.16}_{-0.23}$ & $0.82^{+0.04}_{-0.04}$ & $0.22^{+0.05}_{-0.05}$ & $0.60^{+0.07}_{-0.09}$ &1.01\\
 [2.4pt]
\hline
 Low $z$   & [13.5, 14.5) &  [0.2, 0.3)  &  5464 &$0.45^{+0.16}_{-0.22}$ & $0.94^{+0.06}_{-0.05}$ & $0.47^{+0.11}_{-0.09}$ & $0.60^{+0.03}_{-0.03}$ &1.04 \\
 [2.4pt]
High $z$   & [13.5, 14.5) &  [0.3, 0.5)  &  17834  &$0.46^{+0.18}_{-0.25}$ & $0.90^{+0.05}_{-0.05}$ & $0.33^{+0.06}_{-0.05}$ & $0.57^{+0.03}_{-0.03}$ &1.52\\
 [2.4pt]
\hline
 Low $\lambda$  & [13.5, 14.5) &  [0.2, 0.5)  & 10203 &$0.40^{+0.20}_{-0.25}$ & $0.75^{+0.05}_{-0.04}$ & $0.30^{+0.09}_{-0.06}$ & $0.62^{+0.04}_{-0.04}$ &1.97\\
 [2.4pt]
 High $\lambda$ & [13.5, 14.5) &  [0.2, 0.5) &13095  &$0.63^{+0.18}_{-0.21}$ & $1.02^{+0.07}_{-0.06}$ & $0.38^{+0.07}_{-0.10}$ & $0.51^{+0.04}_{-0.09}$ & 1.60 \\
 [2.4pt]
\hline
\end{tabular}  
\end{table*} 

\subsection{All Samples}\label{sec:all}
\begin{figure}[ht!]
\centering
\includegraphics[width=0.4\textwidth]{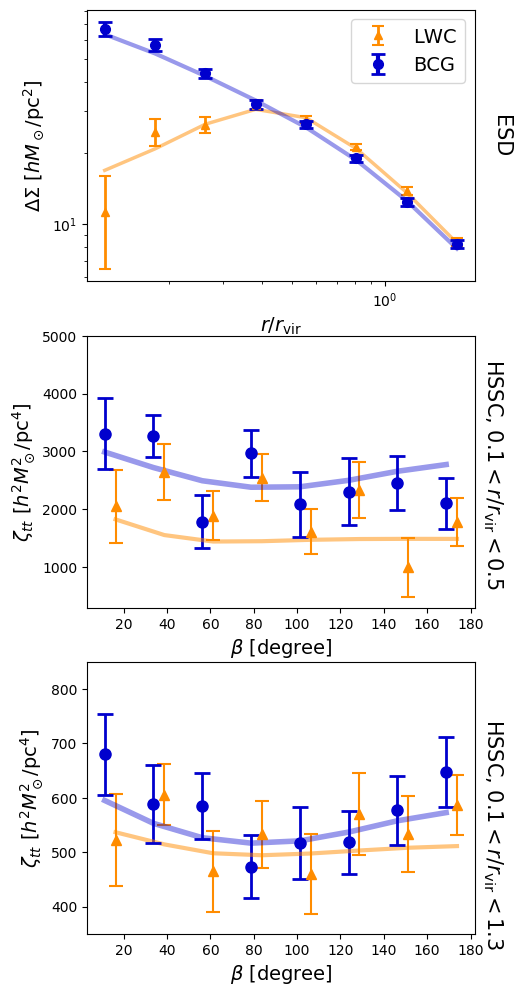}
 
\caption{The measurement results of ESD and HSSCs using BCG (blue) or LWC (orange) as the halo center. The upper panel displays the ESD measurements for all samples, while the middle and the lower panels show the tangential auto-correlated components of HSSC for different maximum radii. The solid lines represent their corresponding best-fit curves. \label{fig:totres}}
\end{figure}

\begin{figure}[ht!]
\centering
\includegraphics[width=0.4\textwidth]{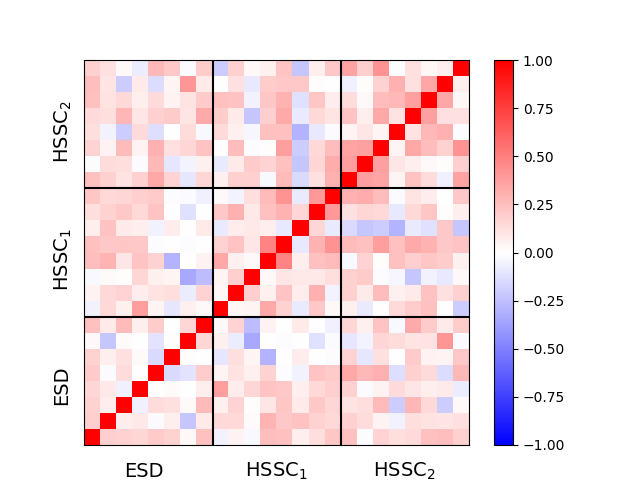}
\caption{Correlation matrix of data points of ESD and HSSC within two different radii for all lens samples. HSSC$_1$ and HSSC$_2$ respectively refer to HSSC results within $0.1 < r_{\rm vir} < 0.5$ and $0.1 < r_{\rm vir} < 1.3$. The corresponding data points are shown in Figure \ref{fig:totres} \label{fig:cov}}
\end{figure}

The upper panel of Figure \ref{fig:totres} shows the result of ESD, and the lower two panels show the tangential auto-correlated components of HSSC within two radius ranges. In each panel, we include the best-fit theoretical curves. The blue points and orange triangles represent outcomes that use BCGs and LWCs, respectively, as the halo center. In this subsection, we focus on analyzing the results centered on BCGs (blue). In the middle and lower panel, the minimum radius is fixed at 0.1$r_{\rm vir}$, and the maximum radii are chosen to be 0.5$r_{\rm vir}$ and 1.3$r_{\rm vir}$ respectively. The average magnitude of HSSC primarily stems from the monopole component, roughly equal to $(\Delta \Sigma)^2$, while its amplitude, which varies in a cosine-like manner, contains information on halo ellipticity. Figure \ref{fig:cov} provides a graphical representation of the correlation matrix of all the data points in Figure \ref{fig:totres}, showing a slight positive correlation among the majority of data. We also calculate the signal-to-noise ratio (S/N) of the anisotropic part of HSSC by using
\begin{equation}
\bigg ( \frac{{\rm S}}{{\rm N}} \bigg )^2 =\sum_{i,j}  (\zeta_{\rm tt}(\beta_i)-\overline{\zeta_{\rm tt}}) C_{i,j}^{-1} (\zeta_{\rm tt}(\beta_j)-\overline{\zeta_{\rm tt}})^\top
\end{equation}
where $C$ is the covariance matrix of HSSCs. The signal $\zeta_{\rm tt}(\beta_i)$ is subtracted by its mean value to isolate the S/N of the anisotropic component. As a result, the S/N is found to be 3.8 for the radial range $0.1 < r/r_{\rm vir} < 0.5$ and 2.8 for $0.1 < r/r_{\rm vir} < 1.3$. While the S/N of HSSC is higher at smaller radii, it is more sensitive to off-centering effects. In contrast, at larger radii, the S/N is lower but contains more information related to ellipticity.

\begin{figure}[ht!]
\centering
\includegraphics[width=0.47\textwidth]{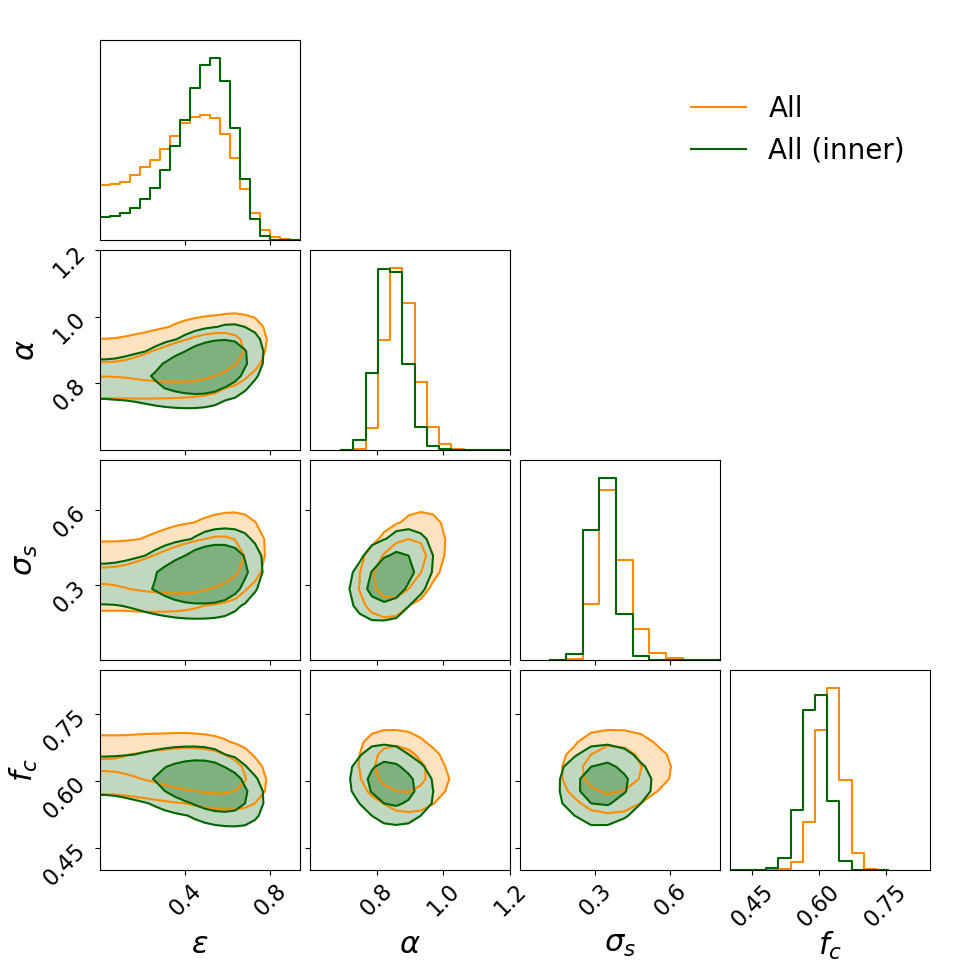}
\caption{The 68\% and 95\% confidence level contour plots of all lens samples. Green contours are obtained from the results of ESD and HSSC within $0.1 < r_{\rm max}/r_{\rm vir} < 0.5$ and $0.1 < r_{\rm max}/r_{\rm vir} < 1.3$, whose data points are shown in Figure \ref{fig:totres}. The results shown in the yellow contours involve only the HSSC within $0.1 < r_{\rm max}/r_{\rm vir} < 0.5$ and the ESD, representing the inner part of halos. \label{fig:contour}}
\end{figure}

The row labeled ``All'' in Table \ref{tab:baseres} displays the parameter constraints for the entire group sample and the corresponding $\chi^2$ per degree of freedom (${\rm d.o.f.} = 12$). Figure \ref{fig:contour} shows the 68\% and 95\% CL contour plots. We find that the halo mass provided in the catalog is biased high, thus yielding an \( \alpha \) value less than 1. For the halo ellipticity, we obtain results that are more than 2$\sigma$ away from zero, and the data from the HSSC clearly exhibit the cosine shape caused by ellipticity, as shown in Figure \ref{fig:totres}. Furthermore, we also use the ESD and inner part HSSC, shown in the upper and middle pannel of Figure \ref{fig:totres}, to obtain the ellipticity within inner radius. The results are shown in Table \ref{tab:baseres} labeled by ``All (inner)'', which shows a slightly smaller ellipticity in inner halos. This is consistent with conclusions from hydrodynamics simulations, which is commonly linked to baryonic cooling leading to more spherically shaped halos \citep{springel2004shapes, kazantzidis2004effect, Cataldi2023}, or possibly to other dark matter models such as SIDM \citep{Peter2013, Despali2022}.

\begin{figure}[ht!]
\centering
\includegraphics[width=0.4\textwidth]{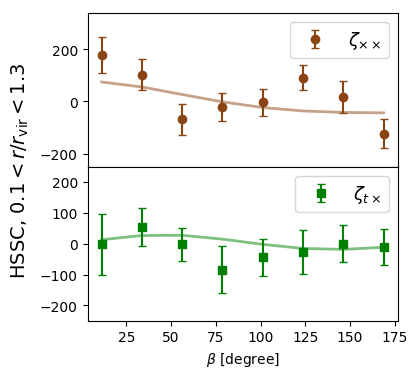}
\caption{HSSC measurements correlated with cross-component within $0.1 < r/r_{\rm vir} < 1.3$. Upper, and lower panels respectively show the results of $\zeta_{\times \times}$, and $\zeta_{t \times}$. The solid curves are the theoretical predictions using the parameters obtained from data points in Figure \ref{fig:totres}.
\label{fig:gxcorrs}}
\end{figure}

Besides, we also show two HSSCs involving the cross components for radius within $0.1 < r/r_{\rm vir} < 1.3$ in Figure \ref{fig:gxcorrs}. Note that they are not used in constraining the parameters due to their low signal-to-noise ratios. The solid curves are the theoretical predictions using the parameters obtained from ESD and tangential auto-correlated HSSC measurements in Figure \ref{fig:totres}. We will discuss the utilization of the cross-components in our future work.

\begin{table*} 
\centering  
\caption{Fitting results for all parameters and corresponding $\chi^2$/d.o.f. under different off-centering models. $q(\varepsilon)$ refers to the model in Eq.\ref{offrthe}, assuming the distribution of off-centering share the same ellipticities as halos. $q=1$ and $q \rightarrow 0$ denote isotropic distribution and BCG alignment along the halo's major axis, respectively.  } \label{tab:diffoff}
\begin{tabular}{l|ccccc}
\hline  
 Eq.\ref{offrthe} &  $\varepsilon (r_{1.3})$ & $\alpha$ & $\sigma_s$ & $f_c$ & $\chi^2$/d.o.f. \\
[2.4pt]
\hline
$q(\varepsilon)$  & $0.48^{+0.12}_{-0.19}$ & $0.84^{+0.04}_{-0.04}$ & $0.33^{+0.05}_{-0.04}$ & $0.59^{+0.03}_{-0.03}$ & 1.03\\
$q=1$ & $0.56^{+0.14}_{-0.20}$ &$0.81^{+0.03}_{-0.03}$ &$0.30^{+0.03}_{-0.03}$ &$0.60^{+0.02}_{-0.02}$ & 0.96\\
$q \rightarrow 0$ &$0.45^{+0.15}_{-0.19}$ & $0.83^{+0.04}_{-0.03}$& $0.35^{+0.05}_{-0.04}$ & $0.57^{+0.03}_{-0.03}$ &  1.05\\
\hline
\end{tabular}  
\end{table*} 

\subsection{Off-centering effect}\label{sec:offcen-result}
We also use LWCs to calculate ESD and HSSC and to fit the halo ellipticity, as shown in Figure \ref{fig:totres}. In the inner radius, the ESD signal is significantly lower, while it is higher in the outer radius, indicating that LWCs suffer from a more severe off-centering effect compared to using BCGs as the halo center. In the HSSC measurements, we find that $\zeta_{\rm tt}$ with LWCs is generally lower, especially at small radii, as tested in the \S\ref{app:off}. Furthermore, the variation of $\zeta_{\rm tt}$ with $\beta$ is weaker than the BCG case (blue points). The fitting results of the parameters are presented in the ``All (LWC)'' row of Table \ref{tab:baseres}. We find that the fraction of well-centered halos is very low, only about 10\%, and the measured ellipticity is significantly lower than the results with BCG center. This indicates that the off-centering effect would diminish the anisotropy of matter distribution and lead to an underestimation of ellipticity, whereas BCGs are relatively better centers. Therefore, we use BCGs as the center in subsequent analyses.

In \S\ref{app:off}, we observe that the spurious HSSC signals caused by off-centers at different locations vary, indicating that the distribution of off-centering may also affect our measurement of ellipticity. In our model, we assume that the anisotropic probability distribution of off-centering anisotropy shares the same projected ellipticity as the halo, as shown in Eq.\ref{offrthe}. To examine the robustness of our assumption, we consider two extreme scenarios: one in which the off-center distribution is isotropic ($q=1$ in Eq.\ref{offrthe}), and the other where the BCGs are aligned perfectly along the major axis of the halo ($q \rightarrow 0$). Table \ref{tab:diffoff} presents all constraints and the $\chi^2$ for all samples within 1.3 virial radius for these three cases. We find that the ellipticity in isotropic case (``$q=1$'') is larger than those in the other two cases. This is because off-center that occurs in the direction perpendicular to the major axis would cause the $\zeta_{\rm tt}(\beta)$ curve to be flatter than when the off-centering is along the major axis, as demonstrated by the green dashed-dotted line and the red dashed line in Figures \ref{fig:off} and \ref{fig:offr} in \S\ref{app:off}. Therefore, in the case of $q=1$, a larger ellipticity is required to compensate for this off-centering effect in order to match the observations. However, the impact of various off-centering models on the ellipticity results falls within the 1$\sigma$ range, and their $\chi^2$ are similar.

\subsection{Dependence on Mass and Redshift}

\begin{figure*}[ht!]
\centering
\includegraphics[width=0.9\textwidth]{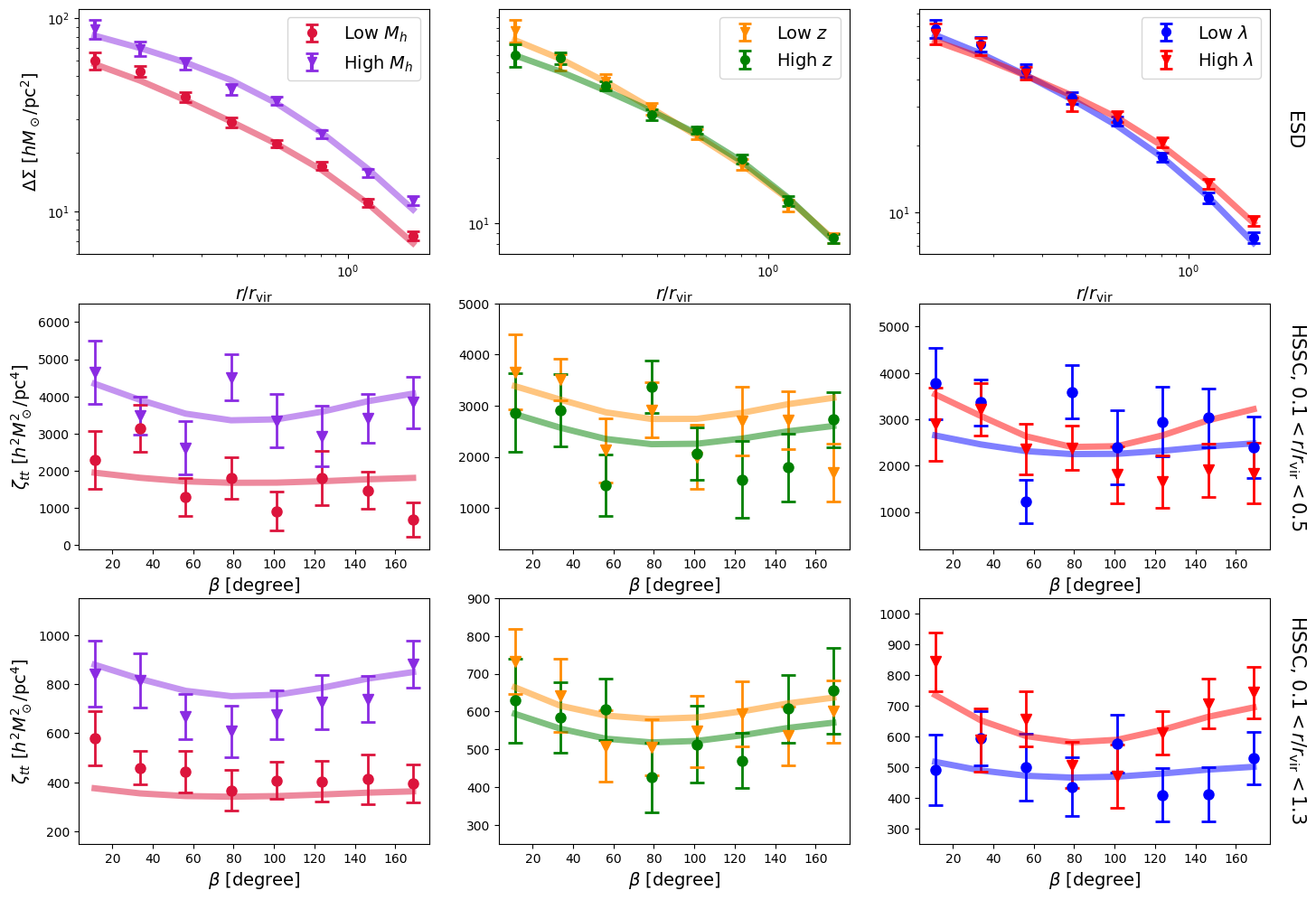}
\caption{The ESD and HSSC measurement results for different subsamples, along with their corresponding best-fit curves. The results in the left, middle, and right panels are derived from lenses of varying masses, redshifts, and richness, respectively. The numerical results for parameters are presented in Table \ref{tab:baseres}. \label{fig:highlow}}
\end{figure*}

We divide the samples into low mass and high mass sub-samples using a halo mass threshold of $10^{14} M_\odot /h$. From the ESD and HSSC results in left panels of Figure \ref{fig:highlow}, we can clearly observe that the lensing signal for the high-mass sample is stronger than that for the low-mass sample, and in all radial ranges, the HSSC measurements for ``High $M_h$'' reveal more pronounced curvature. Table \ref{tab:baseres} details the constraints on their projected ellipticities. We find that lenses with higher mass exhibit higher projected ellipticities compared to lower mass lenses, although the results fall within a 1$\sigma$ consistency range. This could be due to the fact that larger mass halos form later, with ongoing material inflow and mergers at the periphery, making them more elongated. Our findings are consistent with the conclusions of most simulations \citep{jing2002triaxial,Allgood2006,Despali2014, Bonamigo2015}.

To investigate whether the ellipticity of the sample changes with redshift, we also separate the sample into low $z$ and high $z$ groups based on the lens redshift at 0.3. Our results show very similar lensing patterns in the middle panels of Figure \ref{fig:highlow} and halo ellipticities in Table \ref{tab:baseres}. This is reasonable, as the redshifts of the two sub-samples are very close on a cosmic timescale, and their masses are almost identical.

\subsection{Effect of Richness}
\cite{HSSC2017} suggest that the presence of subhalos could lead to an overestimation of HSSC compared with the results from simulation. Here, we delve deeper into the effect of richness. It is generally assumed that a galaxy is enveloped by a halo or subhalo, allowing us to explore the influence of subhalo richess on measurements and ellipticity through the abundance of galaxy clusters. Since the halo mass in the group catalog is given by abundance matching, there is a strong correlation between the mass and richness $\lambda$ in the catalog. Thus, we evenly divided clusters into high- and low-$\lambda$ groups within each narrow redshift and mass bin to make that the 2 samples have similar redshift and group mass distributions. Ultimately, we obtain low-$\lambda$ and high-$\lambda$ samples both with an average mass of $10^{13.7} M_\odot /h$.

The right panel of Figure \ref{fig:highlow} displays the ESDs and HSSCs from halos with different richness. It is evident that the ESDs at small radius are consistent, but in the periphery of high-$\lambda$ lenses is higher than that of low-richness, indicating considerable massive subhalos, which can be described by the off-centering model. Moreover, in the large-radius range, the HSSC of clusters with higher abundances varies more noticeably with $\beta$. Table \ref{tab:baseres} presents all the constraints of parameters for differents samples. \( \alpha \) is highly consistent with 1 for high-$\lambda$ samples, but shows lower halo masses for low-$\lambda$ lenses. For their ellipticity, we find that high-$\lambda$ halos yield larger projected ellipticities, while low-$\lambda$ lenses are more isotropic. Despite taking into account the differences in mass between the two different richness samples, we believe that the difference in their ellipticity also stems from the difference in abundance. In terms of the environment of halos, those of the similar mass but higher richness are dynamically younger and more prone to incorporating peripheral subhalos during material accretion, resulting in higher ellipticities.

In the right middle panel, we notice that the $\zeta_{\rm tt}$ measurements for the ``High-$\lambda$'' samples are lower at large $\beta$, showing a deviation from the theoretical model. Given that the ``High-$\lambda$'' samples contain more satellite galaxies, we speculate that this deviation might be caused by substructures or subhalos within halos. In \S\ref{app:off}, we find that off-centering results in $\zeta_{\rm tt}$ being lower at large $\beta$ compared to small $\beta$. We could liken subhalos to small off-centered halos, which might induce similar systematic biases. Nonetheless, modeling the substructure of halos is complex, and numerical simulations or more accurate models could be used in the future for more detailed modeling.

As we define halo richness based on the number of member galaxies in clusters, we evaluate the projected ellipticity of satellite galaxies to check for similar properties. The projected ellipticity of satellite galaxies can be derived from their quadrupole moments, with 
\begin{equation}
\varepsilon_1=\frac{Q_{11}-Q_{22}}{Q_{11}+Q_{22}},
\end{equation}
\begin{equation}
\varepsilon_2=\frac{2 Q_{12}}{Q_{11}+Q_{22}}
\end{equation}
and 
\begin{equation}\label{eq:esat}
\varepsilon_{\rm sat} = \sqrt{\varepsilon_1^2 + \varepsilon_2^2}.
\end{equation}
The quadrupole moment is defined in Eq.\ref{Qij} but with the respect to the BCG center. Due to the large uncertainties in ellipticity from the clusters with low richness and the potential for Poisson sampling bias to increase their projected ellipticity, we exclude galaxy clusters with fewer than 20 members from our sample. 
We find that the ellipticities of satellite galaxy distribution in low-$\lambda$ and high-$\lambda$ groups are very similar, being $0.351 \pm 0.037$ and $0.353 \pm 0.037$, respectively. This result contrasts with our findings from lensing measurements (in Table \ref{tab:baseres}) and simulations (in \S\ref{sec:siml}), which show ellipticity increases with $\lambda$. Although these trends are within 1$\sigma$ range, the ellipticities of satellite galaxy distribution do not exhibit a similar variation. This discrepancy may be attributed to the positional distribution of galaxies not effectively tracing dark matter density, or it could be due to other systematic errors, such as the inclusion of randomly distributed interlopers in the satellite sample, which would reduce the ellipticity \citep{gonzalez2021measuring}.

\subsection{Ellipticity in IllustrisTNG}\label{sec:siml}
\begin{table} 
\centering  
\caption{The ellipticity of dark matter halos in the TNG100-1 simulation. The ``Low $z$'' samples are selected from the snapshot at $z=0.24$, the ``High $z$'' samples are from the snapshot at $z=0.4$, and the remaining samples are from the snapshot at $z=0.3$. The selection criteria for halo mass $M_h$ and richness $\lambda$ are consistent with those in lensing measurements in Table \ref{tab:baseres}.} \label{tab:simu}
\begin{tabular}{l|cc}
\hline  
   & $\varepsilon_{\rm dm}$& $\varepsilon_{\rm sat}$ \\
[2.4pt]
\hline
All   & $0.32 \pm 0.17$ & $0.29 \pm 0.17$\\
Low $z$ & $0.32 \pm 0.16$  & $0.28 \pm 0.17$\\
High $z$ & $0.28 \pm 0.14$  & $0.26 \pm 0.14$\\
Low $M_h$ & $0.33 \pm 0.19$ & $0.31 \pm 0.18$\\
High $M_h$ &  $0.27 \pm 0.11$ & $0.23 \pm 0.12$ \\
Low $\lambda$ & $0.23 \pm 0.11$  & $0.21 \pm 0.09$\\
High $\lambda$ & $0.42 \pm 0.20$  & $0.38 \pm 0.19$\\
\hline
\end{tabular}  
\end{table} 

To deepen our understanding and validate our findings, we calculate the projected ellipticity of halos under similar conditions in the IllustrisTNG simulation\footnote{https://www.tng-project.org/} \citep{illutrisTNGNelson}. IllustrisTNG is an advanced cosmological hydrodynamic simulation that enables detailed modeling of dark matter halo formation and evolution, while also accounting for the complex effects of baryons. We utilize the dark matter halo catalog identified by the FOF algorithm for the TNG100-1 simulation, which provides a balanced combination of a large box size of 110.7 Mpc and a high mass resolution of $7.5 \times 10^6 {\rm M_\odot}$ in the suite. We identify dark matter halos with mass $M_{\rm 200 m}> 10^{13.5} M_\odot /h$ from the snapshot at redshifts of 0.3 (for ``All''), 0.24  (for ``Low $z$'') and 0.4 (for ``High $z$''). We then use the mass weighted center of halos (labeled by ``GroupCM'' in catalog) and calculate the projected ellipticities of darkmatter and satellite galaxies using Eq.\ref{eq:esat}. 

The outcomes are summarized in Table \ref{tab:simu}. We find that although the projected ellipticity of satellite systems is slightly smaller than that of dark matter halos, they are still very consistent. Additionally, there is no significant difference in the ellipticities of halos at different redshifts, consistent with our lensing results. 

\begin{figure}[ht!]
\centering
\includegraphics[width=0.4\textwidth]{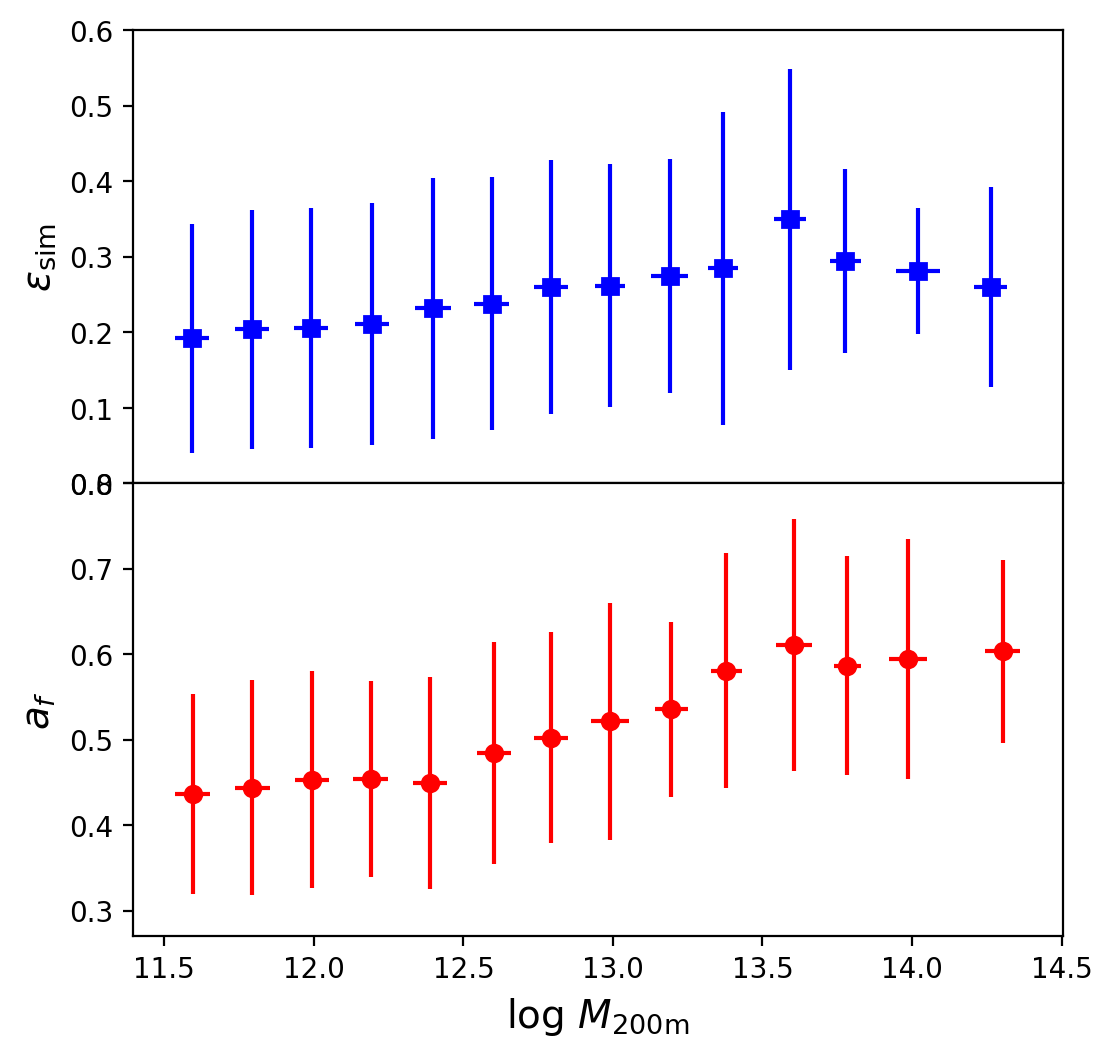}
\caption{The projected ellipticity $\varepsilon_{\rm sim}$ of dark matter halos of different masses and their formation times $a_f$ in the IllustrisTNG simulation. \label{fig:Mellaf}}
\end{figure}

We divide the halos in the ``All'' sample into low-mass and high-mass groups based on the threshold of $M_{\rm 200 m}=10^{14} M_\odot / h$, whose results are presented in Table \ref{tab:simu}. Surprisingly, the results show that high-mass halos exibit smaller ellipticities, which contradicts the conclusions from other simulation studies. To further investigate, we expand the mass range of halos to include those with $M_{\rm 200 m}> 10^{11.5} M_\odot /h$ for ellipticity calculation and shown them in Figure \ref{fig:Mellaf}. We observe that the projected ellipticities primarily increased with halo mass in the low and medium mass range, but reversed trend for halos with mass larger than $10^{13.5} M_\odot /h$. Subsequently, we refer to the halo structure catalog by \cite{Anbajagane2022}, examining the relationship between halo formation time and their mass. It is found that, typically, the formation time becomes later as mass increases, but when the halo mass exceeds $10^{13.5} M_\odot /h$, the formation time remains about $a_f \sim 0.6$. This suggests that these massive halos have a similar evolutionary period post-formation, and more massive halos can gather more quickly and weaken their connection to the cosmic web, leading to a more spherical shape.

Finally, we divide halos in the ``All'' sample into low-richness and high-richness groups, ensuring the identical mass and redshift distributions as the same way as we do in the lensing measurements. We find that halos with same mass but higher abundance exhibit larger ellipticities in Table \ref{tab:simu}, aligning with our measurement outcomes. The formation times $a_f$ for low-$\lambda$ and high-$\lambda$ halos were calculated to be 0.54 and 0.66, respectively. This suggests that high-$\lambda$ halos form later and are more likely to be in an accretion stage, hence the higher ellipticities. This further validates our explanation for the lensing results. Moreover, this finding is inconsistent with outcomes from the ellipticities of satellite galaxy distribution, suggesting that galaxy clusters may not effectively track the properties of halos in this regard.

\section{Conclusion}\label{sec:conclusion}

Our work is the first to use halo-shear-shear correlations (HSSC) to measure the ellipticity of galaxy cluster halos, avoiding the conventional challenges of determining the major axis in galaxy-galaxy lensing measurements. We consider the off-centering effects in our model of HSSC, which has slight degeneracy with halo ellipticity. We break this degeneracy by jointly constraining the monopole ESD and high-order HSSC, thereby deriving the halo's projected ellipticity. The galaxy cluster catalog is sourced from DESI, and we select the clusters with halo masses spanning from $10^{13.5}$ to $10^{14.5} M_\odot / h$ in the redshift range of [0.2, 0.5]. The shear catalog is derived from the HSC processed by the FQ pipeline, covering an extensive area of 1400 deg$^2$ and comprising approximately 100 million galaxies. We employ the PDF-SYM method for measuring ESD and HSSC, enhancing statistical information and reducing statistic bias. We explore how the halo ellipticity varies with radius, halo mass, redshift, and the richness of galaxy clusters. Despite loose constraints on ellipticity, we still identify its variation trends with these properties. Finally, we use the IllustrisTNG simulation to verify our results. Our findings are concluded as follows:

\begin{itemize}
\item For all samples, we measure their average projected ellipticity within 1.3 varial radius to be $0.48^{+0.12}_{-0.19}$. We find that the ellipticity of dark matter halos tends to be larger in the outskirts compared to the inner regions. 

\item We divide the sample into high- and low-mass subgroups and find that high-mass halos exhibit larger ellipticities. Utilizing dark matter halo catalogs from the IllustrisTNG simulation, it is observed that ellipticities increase with mass in the low to medium range, but decline for masses above $10^{13.5} M_\odot / h$. These trends can be ascribed to the formation times of halos.

\item We observe that halos at different redshifts show very similar ellipticities.

\item We conclude that high-richness halos with same masses exhibit larger ellipticities, indicating that these halos form later and are in a more active stage of mass accretion. This also implies that the richness of a halo can be an indicator of its dynamical age and the extent of its interaction with the cosmic web. 
\end{itemize}

Our research provides valuable insights into the formation and structural evolution of dark matter halos, which is essential for testing and refining cosmological models. Moreover, the methodology established in this work can serve as a foundation for future investigations, encouraging the development of more sophisticated techniques to further unravel the mysteries of halos and cosmos.

\textbf{Acknowledgements} Thanks for the valuable suggestions and insightful discussions provided by Wenting Wang, Xiaolin Luo and Zhiwei Shao. This work is supported by the National Key Basic Research and Development Program of China (2023YFA1607800, 2023YFA1607802), the NSFC grants (11621303, 11890691, 12073017), and the science research grants from China Manned Space Project (No. CMS-CSST-2021-A01). The computations in this paper were run on the $\pi$ 2.0 cluster supported by the Center of High Performance Computing at Shanghai Jiaotong University, and the Gravity supercomputer of
the Astronomy Department, Shanghai Jiaotong University.

\appendix


\section{The impacts of off-centering on HSSCs}\label{app:off}
\begin{figure*}[ht!]
\centering
\includegraphics[width=0.8\textwidth]{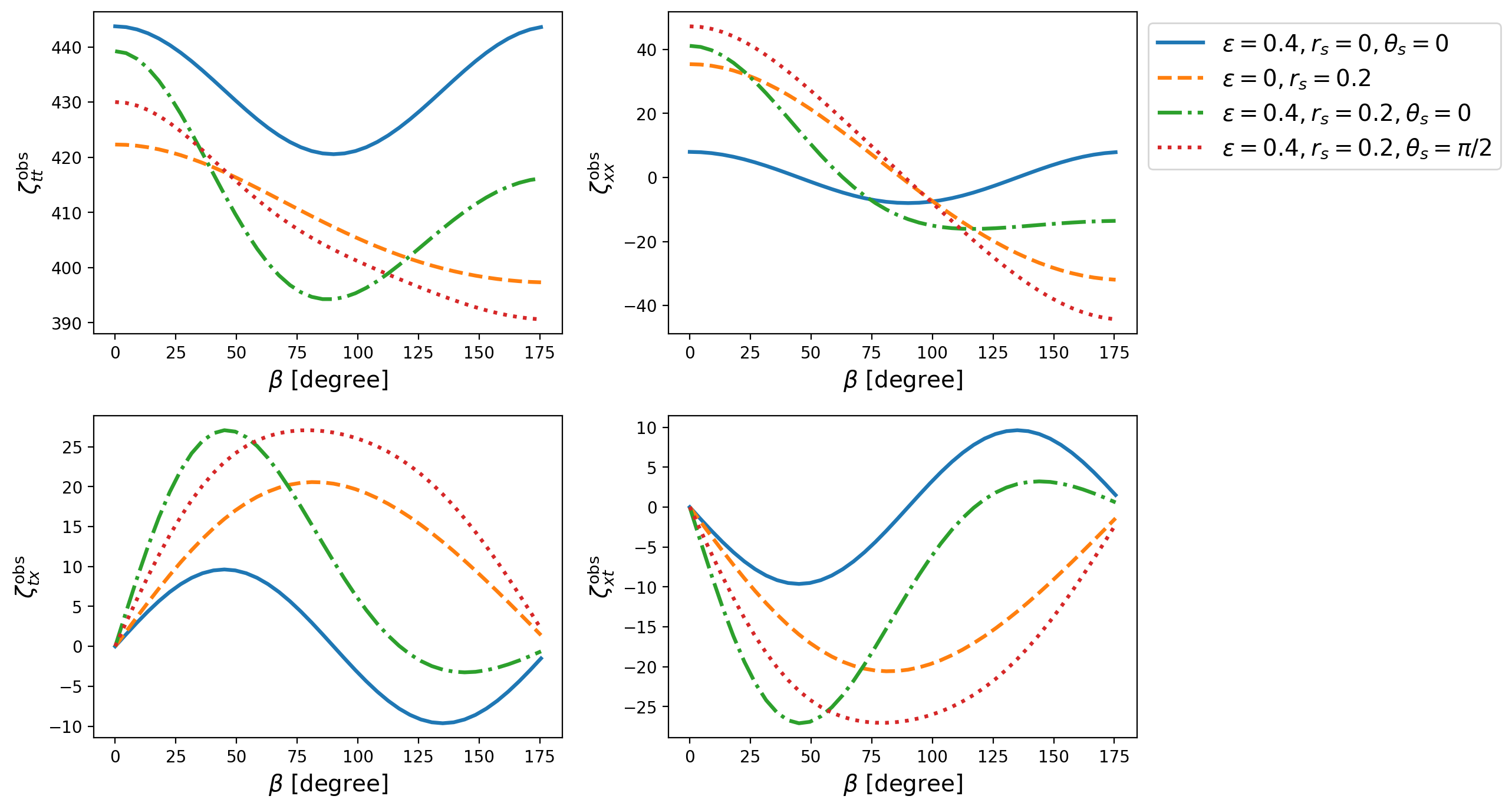}
\caption{Four components of HSSCs with different off-centering effects. The halo mass is set to $10^{14} M_\odot / h$ and located at $z = 0.3$. The blue solid line represents HSSCs with an ellipticity of 0.4 and no off-centering effects. The orange dashed line represents HSSCs with an ellipticity of 0 and a central offset at 0.2 $r_{\rm vir}$. The green dot-dashed line and red dotted line correspond to cases where the offset center is located along the major axis of halo and perpendicular to the major axis, respectively.} \label{fig:off}
\end{figure*}

\begin{figure*}[ht!]
\centering
\includegraphics[width=0.8\textwidth]{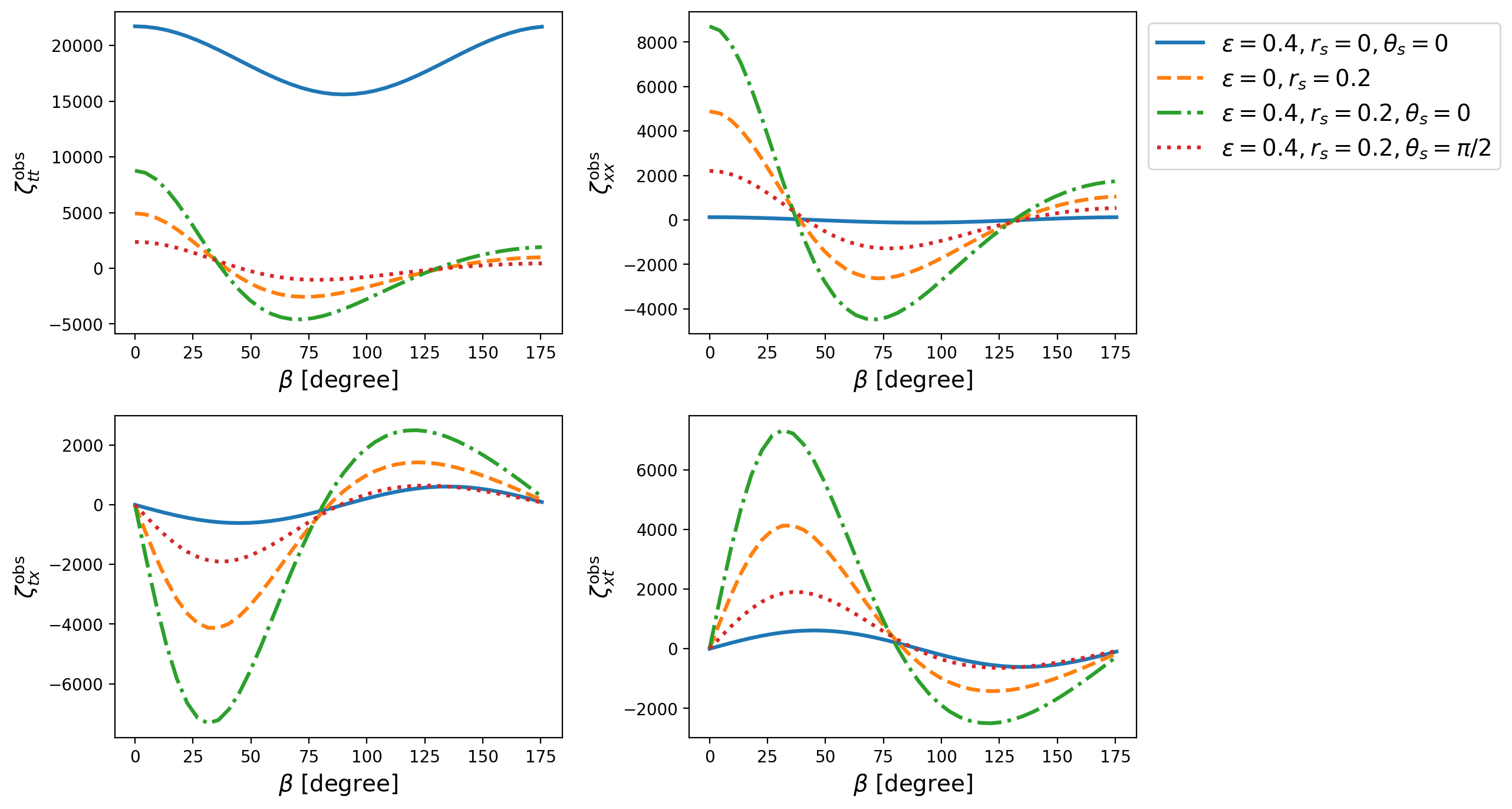}
\caption{Similar to Figure \ref{fig:off}, but the radii of shears are both at 0.1 viral radius.} \label{fig:offr}
\end{figure*}

In this section, we discuss the impact of off-centering in the theoretical model for a single halo on the components of the HSSCs. Taking a halo with a projected ellipticity of $\varepsilon =0.4$, a mass of $10^{14} M_\odot / h$ and located at $z=0.3$ as an example, the blue solid line in the Figure \ref{fig:off} illustrates the four HSSC components at 1 virial radius without off-centering effect. Clear signals varying with $\beta$, induced by the ellipticity, can be observed, as described in Eqs.\ref{xittcen} and \ref{xixxcen}. However, when the ellipticity is set to 0 and the center is offset by 0.2 virial radius, the orange dashed lines in the figure illustrate the corresponding HSSCs using Eq.\ref{eq:obsij}. We find that off-centering also induces signals that vary with $\beta$, but these signals differ from those caused by ellipticity. Specifically, off-centering results in lower values of $\zeta_{\rm tt}$ and $\zeta_{\times \times}$ at large $\beta$ and suppresses the overall amplitude of $\zeta_{\rm tt}$. For halos with non-zero ellipticity, the impact of off-centering varies with its direction with the respect to the true center. For example, the green dot-dashed lines and red dotted lines in Figure \ref{fig:off} represent the results of offsets along the major axis and perpendicular to the major axis, respectively. We observe that off-centering perpendicular to the major axis has a more significant impact on the HSSC than along the major axis. Additionally, we plot the HSSC components at a small radius of 0.1 virial radii, as shown in the Figure \ref{fig:offr}. At smaller radii, off-centering has a more pronounced effect in suppressing $\zeta_{\rm tt}$ and magnifying other components. Additionally, we note that at different radii, $\zeta_{t \times}$ and $\zeta_{\times t}$ exhibit opposite trends, which can be attributed to the fact that $\gamma_\times$ is negative at smaller radii but positive at larger radii.

\section{Onsite Shear bias calibration using Field Distortion}\label{sec:fd}


Shear biases are typically quantified linearly as:
\begin{equation}\label{eq:gf}
    g_i^{\rm obs}=(1+m_i)g_i^{\rm true}+c_i, 
\end{equation}
in which i=1 or 2, and $m_i$ and $c_i$ are the multiplicative and addictive biases respectively. \cite{Zhang2019ApJ} proposed using the intrinsic distortions of the CCD focal plane (field distortions, FD) to detect $m$ and $c$. FD refers to the image distortion caused by imperfections in the optical system or imaging equipment. These distortions can manifest as twisting, stretching, or compression of the image. This method of calibration uses the properties of the data itself, without the need for simulations or any external datasets. The FQ shear catalog provides two components of FD for each galaxy, $g_{\rm f1}$ and $g_{\rm f2}$. Typically, the average cosmological shear value for all galaxies at the same focal plane position is zero, as they usually cover a large area of the sky. Therefore, by comparing the measured field distortion signals with their true signals in catalog, the biases in shear measurement can be determined.

\begin{figure*}[ht!]
\centering
\includegraphics[width=0.6\textwidth]{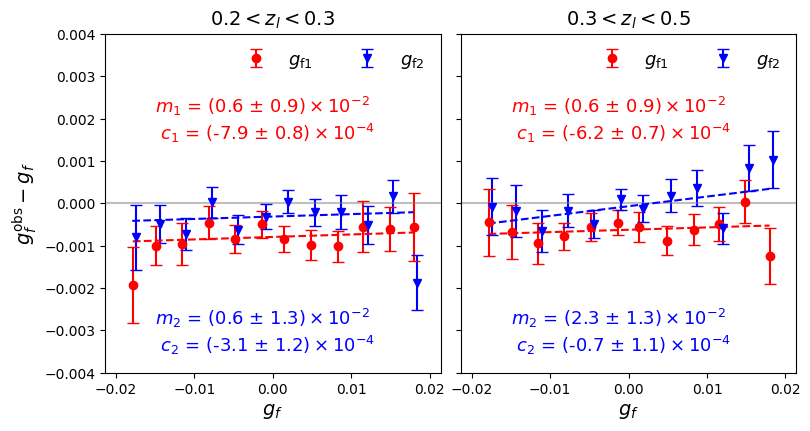}
\caption{The field distortion test for background galaxies of galaxy clusters at two different lens redshift bins. The red and blue data points represent the two shear components \( g_{\rm f1}^{\rm obs} \) or \( g_{\rm f2}^{\rm obs} \). To display them more clearly, the blue points have been horizontally shifted to the right by 0.0005. The dashed lines indicate the fitting lines of Eq. \ref{eq:gf}, and \(m_i\) and \( c_i \) are their fitting results. \label{fig:fd}}
\end{figure*}

The FD test results for the shear catalogs of different bands of HSC are presented in \cite{Liu2024zlh}. It is however necessary to carry such tests again due to the redshift selection/binning of the background galaxies, as shown in \cite{Shen2024}. The resulting shear bias must be caused by selection effects related to the photometric redshift, but the exact physical reason is not clear yet. A convenient feature of the FD test is that one can always carry out such a test for any sub-sample of the shear catalog, which may be subject to selection effects of whatever origin. 

We divide all lens galaxy clusters into two subsamples based on redshift in this test. We then find the background galaxies within a radius range of 0.1-2$r_{\rm vir}$ around the lenses (i.e., all background galaxies measured in the ESD at all radii) and calculate their field distortions. Figure \ref{fig:fd} shows the FD results for the two different redshift subsamples. The vertical axis represents the two observed distortion components \( g_{\rm f1}^{\rm obs} \) (red circles) or \( g_{\rm f2}^{\rm obs} \) (blue triangles) measured by the PDF method, minus their corresponding true values \( g_{\rm f} \). To better approximate reality, each background galaxy is assigned a weight $\Sigma_c$ defined in Eq.\ref{sigmac}. If there were no bias in the shear measurement, the observed values should equal the true values, indicated by the gray line in the figures. The dashed lines in figures show the best-fit results of Eq.\ref{eq:gf}.
It is found that most background galaxy samples have a small multiplicative bias, generally at a few percent level. Therefore, we do not apply any corrections to our shear data.

\bibliography{sample631}{}
\bibliographystyle{aasjournal}



\end{document}